\magnification=\magstep1
\hsize 14cm
\hoffset -0.3cm
\baselineskip=16pt
\font\sm=cmr8 
\catcode`@=11 
%
%
\def\b@lank{ }

\newif\if@simboli
\newif\if@riferimenti
\newif\if@bozze
\newif\if@data

\def\bozze{\@bozzetrue 
\immediate\write16{!!!   INSERISCE NOME EQUAZIONI   !!!}}

\newwrite\file@simboli
\def\simboli{

    \immediate\write16{ !!! Genera il file \jobname.SMB }
    \@simbolitrue\immediate\openout\file@simboli=\jobname.smb
     \immediate\write\file@simboli{Simboli di \jobname}}

\newwrite\file@ausiliario
\def\riferimentifuturi{
    \immediate\write16{ !!! Genera il file \jobname.aux }
    \@riferimentitrue\openin1 \jobname.aux
    \ifeof1\relax\else\closein1\relax\input\jobname.aux\fi
    \immediate\openout\file@ausiliario=\jobname.aux}

\newcount\eq@num\global\eq@num=0
\newcount\sect@num\global\sect@num=0
\newcount\para@num\global\para@num=0
\newcount\const@num\global\const@num=0
\newcount\lemm@num\global\lemm@num=0

\newif\if@ndoppia
\def\numerazionedoppia{\@ndoppiatrue\gdef\la@sezionecorrente{\the\sect@num}}

\def\se@indefinito#1{\expandafter\ifx\csname#1\endcsname\relax}
\def\spo@glia#1>{} 

\newif\if@primasezione
\@primasezionetrue

\def\s@ection#1\par{\immediate
    \write16{#1}\if@primasezione\global\@primasezionefalse\else\goodbreak
    \vskip\spaziosoprasez\fi\noindent
    {\bf#1}\nobreak\vskip\spaziosottosez\nobreak\noindent}


\newif\if@indice
\newif\if@ceindice

\newwrite\file@indice

\def\indice{
           \immediate\write16{Genera il file \jobname.ind}
           \@indicetrue
           \immediate\openin2 \jobname.ind
           \ifeof2\relax\else
             \closein2\relax
       \@ceindicetrue\fi
            \if@ceindice\relax\else
            \immediate\openout\file@indice=\jobname.ind
         \immediate\write
         \file@indice{\string\vskip5pt
         \string{ \string\bf \string\centerline\string{ Indice 
         \string}\string}\string\par}
            \fi
            }

\def\quiindice{\if@ceindice\vfill\eject\input\jobname.ind\else\vfill\eject
       \immediate\write\file@indice{\string{\string\bf\string~ 
       Indice\string}\string\hfill\folio}
       \null\vfill\eject\null\vfill\eject\relax\fi}

%

\def\sezpreset#1{\global\sect@num=#1
    \immediate\write16{ !!! sez-preset = #1 }   }

\def\spaziosoprasez{50pt plus 60pt}
\def\spaziosottosez{15pt}

\def\sref#1{\se@indefinito{@s@#1}\immediate\write16{ ??? \string\sref{#1}
    non definita !!!}
    \expandafter\xdef\csname @s@#1\endcsname{??}\fi\csname @s@#1\endcsname}

\def\autosez#1#2\par{
    \global\advance\sect@num by 1\if@ndoppia\global\eq@num=0\fi
    \global\lemm@num=0
    \global\para@num=0
    \xdef\la@sezionecorrente{\the\sect@num}
    \def\usa@getta{1}\se@indefinito{@s@#1}\def\usa@getta{2}\fi
    \expandafter\ifx\csname @s@#1\endcsname\la@sezionecorrente\def
    \usa@getta{2}\fi
    \ifodd\usa@getta\immediate\write16
      { ??? possibili riferimenti errati a \string\sref{#1} !!!}\fi
    \expandafter\xdef\csname @s@#1\endcsname{\la@sezionecorrente}
    \immediate\write16{\la@sezionecorrente. #2}
    \if@simboli
      \immediate\write\file@simboli{ }\immediate\write\file@simboli{ }
      \immediate\write\file@simboli{  Sezione 
                                  \la@sezionecorrente :   sref.   #1}
      \immediate\write\file@simboli{ } \fi
    \if@riferimenti
      \immediate\write\file@ausiliario{\string\expandafter\string\edef
      \string\csname\b@lank @s@#1\string\endcsname{\la@sezionecorrente}}\fi
    \goodbreak\vskip 48pt plus 60pt
    \noindent{\bf\the\sect@num.\quad #2}
   \if@bozze
    {\tt #1}\fi
    \par\nobreak\vskip 15pt
    \nobreak}

\def\blankii{\blank\blank}

\def\destra#1{{\hfill#1}}
\font\titfnt=cmssbx10 scaled \magstep2
\font\capfnt=cmss17 scaled \magstep4
\def\blank{\vskip 12pt}

\def\capitolo#1#2\par{
    \global\advance\sect@num by 1\if@ndoppia\global\eq@num=0\fi
    \global\lemm@num=0
    \global\para@num=0
    \xdef\la@sezionecorrente{\the\sect@num}
    \def\usa@getta{1}\se@indefinito{@s@#1}\def\usa@getta{2}\fi
    \expandafter\ifx\csname @s@#1\endcsname\la@sezionecorrente\def
    \usa@getta{2}\fi
    \ifodd\usa@getta\immediate\write16
      { ??? possibili riferimenti errati a \string\sref{#1} !!!}\fi
    \expandafter\xdef\csname @s@#1\endcsname{\la@sezionecorrente}
    \immediate\write16{\la@sezionecorrente. #2}
   \if@simboli
      \immediate\write\file@simboli{ }\immediate\write\file@simboli{ }
      \immediate\write\file@simboli{  Sezione 
                                  \la@sezionecorrente :   sref.   #1}
      \immediate\write\file@simboli{ } \fi
    \if@riferimenti
      \immediate\write\file@ausiliario{\string\expandafter\string\edef
      \string\csname\b@lank @s@#1\string\endcsname{\la@sezionecorrente}}\fi
           \par\vfill\eject
           \destra{\capfnt {\la@sezionecorrente}\hbox to 10pt{\hfil}}
           \blankii\noindent{\titfnt\baselineskip=20pt
           \hfill\uppercase{#2}}\blankii
      \if@indice
       \if@ceindice\relax\else\immediate\write
       \file@indice{\string\vskip5pt\string{\string\bf 
       \la@sezionecorrente.#2\string}\string\hfill\folio\string\par}\fi\fi
       \if@bozze
         {\tt #1}\par\fi\nobreak}

\def\semiautosez#1#2\par{
    
\gdef\la@sezionecorrente{#1}\if@ndoppia\global\eq@num=0
     \fi
     \global\lemm@num=0
    \global\para@num=0
        \if@simboli
      \immediate\write\file@simboli{ }\immediate\write\file@simboli{ }
      \immediate\write\file@simboli{  Sezione ** : sref.
          \expandafter\spo@glia\meaning\la@sezionecorrente}
      \immediate\write\file@simboli{ }\fi
    \s@ection#2\par}


\def\pararef#1{\se@indefinito{@ap@#1}
    \immediate\write16{??? \string\pararef{#1} non definito !!!}
    \expandafter\xdef\csname @ap@#1\endcsname {#1}
    \fi\csname @ap@#1\endcsname}

\def\autopara#1#2\par{
     \global\advance\para@num by 1
     \xdef\il@paragrafo{\la@sezionecorrente.\the\para@num}
     \vskip10pt
     \noindent {\bf \il@paragrafo\ #2}
     \def\usa@getta{1}\se@indefinito{@ap@#1}\def\usa@getta{2}\fi
     \expandafter\ifx\csname  @ap@#1\endcsname\il@paragrafo\def\usa@getta{2}\fi
     \ifodd\usa@getta\immediate\write16
        {??? possibili riferimenti errati a \string\pararef{#1} !!!}\fi
     \expandafter\xdef\csname @ap@#1\endcsname{\il@paragrafo}
     \def\usa@getta{\expandafter\spo@glia\meaning
     \la@sezionecorrente.\the\para@num}
     \if@simboli
      \immediate\write\file@simboli{ }\immediate\write\file@simboli{ }
      \immediate\write\file@simboli{ paragrafo
                                  \il@paragrafo :   pararef.   #1}
      \immediate\write\file@simboli{ } \fi
    \if@riferimenti
      \immediate\write\file@ausiliario{\string\expandafter\string\edef
      \string\csname\b@lank @ap@#1\string\endcsname{\il@paragrafo}}\fi
    \if@indice
     \if@ceindice\relax\else\immediate\write
       \file@indice{\string\noindent\string\item\string{
       \il@paragrafo.\string}#2\string\dotfill\folio\string\par}\fi\fi
    \if@bozze
       {\tt #1}\fi\par\nobreak\vskip .3 cm \nobreak}


\def\eqpreset#1{\global\eq@num=#1
     \immediate\write16{ !!! eq-preset = #1 }     }

\def\eqlabel#1{\global\advance\eq@num by 1
    \if@ndoppia\xdef\il@numero{\la@sezionecorrente.\the\eq@num}
       \else\xdef\il@numero{\the\eq@num}\fi
    \def\usa@getta{1}\se@indefinito{@eq@#1}\def\usa@getta{2}\fi
    \expandafter\ifx\csname @eq@#1\endcsname\il@numero\def\usa@getta{2}\fi
    \ifodd\usa@getta\immediate\write16
       { ??? possibili riferimenti errati a \string\eqref{#1} !!!}\fi
    \expandafter\xdef\csname @eq@#1\endcsname{\il@numero}
    \if@ndoppia
       \def\usa@getta{\expandafter\spo@glia\meaning
       \il@numero}
       \else\def\usa@getta{\il@numero}\fi
    \if@simboli
       \immediate\write\file@simboli{  Equazione 
            \usa@getta :  eqref.   #1}\fi
    \if@riferimenti
       \immediate\write\file@ausiliario{\string\expandafter\string\edef
       \string\csname\b@lank @eq@#1\string\endcsname{\usa@getta}}\fi}

\def\eqsref#1{\se@indefinito{@eq@#1}
    \immediate\write16{ ??? \string\eqref{#1} non definita !!!}
    \if@riferimenti\relax
    \else\eqlabel{#1} ???\fi
    \fi\csname @eq@#1\endcsname }

\def\autoeqno#1{\eqlabel{#1}\eqno(\csname @eq@#1\endcsname)\if@bozze
        {\tt #1}\else\relax\fi}
\def\autoleqno#1{\eqlabel{#1}\leqno(\csname @eq@#1\endcsname)}
\def\eqref#1{(\eqsref{#1})}


\def\lemmalabel#1{\global\advance\lemm@num by 1
    \xdef\il@lemma{\la@sezionecorrente.\the\lemm@num}
    \def\usa@getta{1}\se@indefinito{@lm@#1}\def\usa@getta{2}\fi
    \expandafter\ifx\csname @lm@#1\endcsname\il@lemma\def\usa@getta{2}\fi
    \ifodd\usa@getta\immediate\write16
       { ??? possibili riferimenti errati a \string\lemmaref{#1} !!!}\fi
    \expandafter\xdef\csname @lm@#1\endcsname{\il@lemma}
    \def\usa@getta{\expandafter\spo@glia\meaning
       \la@sezionecorrente.\the\lemm@num}
       \if@simboli
       \immediate\write\file@simboli{  Lemma
            \usa@getta :  lemmaref #1}\fi
    \if@riferimenti
       \immediate\write\file@ausiliario{\string\expandafter\string\edef
       \string\csname\b@lank @lm@#1\string\endcsname{\usa@getta}}\fi}

\def\autolemma#1{\lemmalabel{#1}\csname @lm@#1\endcsname\if@bozze
    {\tt #1}\else\relax\fi}   

\def\lemmaref#1{\se@indefinito{@lm@#1}
    \immediate\write16{ ??? \string\lemmaref{#1} non definita !!!}
    \if@riferimenti\else
    \lemmalabel{#1}???\fi
    \fi\csname @lm@#1\endcsname}


\newcount\cit@num\global\cit@num=0

\newwrite\file@bibliografia
\newif\if@bibliografia
\@bibliografiafalse
\newif\if@corsivo
\@corsivofalse

\def\title#1{{\it #1}}

\def\lp@cite{[}
\def\rp@cite{]}
\def\trap@cite#1{\lp@cite #1\rp@cite}
\def\lp@bibl{[}
\def\rp@bibl{]}
\def\trap@bibl#1{\lp@bibl #1\rp@bibl}

\def\refe@renza#1{\if@bibliografia\immediate        
    \write\file@bibliografia{
    \string\item{\trap@bibl{\cref{#1}}}\string
    \bibl@ref{#1}\string\bibl@skip}\fi}

\def\ref@ridefinita#1{\if@bibliografia\immediate\write\file@bibliografia{ 
    \string\item{?? \trap@bibl{\cref{#1}}} ??? tentativo di ridefinire la 
      citazione #1 !!! \string\bibl@skip}\fi}

\def\bibl@ref#1{\se@indefinito{@ref@#1}\immediate
    \write16{ ??? biblitem #1 indefinito !!!}\expandafter\xdef
    \csname @ref@#1\endcsname{ ??}\fi\csname @ref@#1\endcsname}

\def\c@label#1{\global\advance\cit@num by 1\xdef            
   \la@citazione{\the\cit@num}\expandafter
   \xdef\csname @c@#1\endcsname{\la@citazione}}

\def\bibl@skip{\vskip 5truept}


\def\stileincite#1#2{\global\def\lp@cite{#1}\global
    \def\rp@cite{#2}}
\def\stileinbibl#1#2{\global\def\lp@bibl{#1}\global
    \def\rp@bibl{#2}}

\def\corsivo{\global\@corsivotrue}

\def\citpreset#1{\global\cit@num=#1
    \immediate\write16{ !!! cit-preset = #1 }    }

\def\autobibliografia{\global\@bibliografiatrue\immediate
    \write16{ !!! Genera il file \jobname.BIB}\immediate
    \openout\file@bibliografia=\jobname.bib}

\def\cref#1{\se@indefinito                  
   {@c@#1}\c@label{#1}\refe@renza{#1}\fi\csname @c@#1\endcsname}

\def\cite#1{\trap@cite{\cref{#1}}}                  
\def\ccite#1#2{\trap@cite{\cref{#1},\cref{#2}}}     
\def\ncite#1#2{\trap@cite{\cref{#1}--\cref{#2}}}    
\def\upcite#1{$^{\,\trap@cite{\cref{#1}}}$}               
\def\upccite#1#2{$^{\,\trap@cite{\cref{#1},\cref{#2}}}$}  
\def\upncite#1#2{$^{\,\trap@cite{\cref{#1}-\cref{#2}}}$}  

\def\clabel#1{\se@indefinito{@c@#1}\c@label           
    {#1}\refe@renza{#1}\else\c@label{#1}\ref@ridefinita{#1}\fi}


\def\biblskip#1{\def\bibl@skip{\vskip #1}}           

\def\insertbibliografia{\if@bibliografia             
    \immediate\write\file@bibliografia{ }
    \immediate\closeout\file@bibliografia
   \if@indice
     \if@ceindice\relax\else\immediate\write
       \file@indice{\string\vskip5pt\string{\string\bf\string~ 
       Bibliografia\string}\string\hfill\folio\string\par}\fi\fi
     \catcode`@=11\input\jobname.bib\catcode`@=12\fi
   }


\def\commento#1{\relax} 
\def\biblitem#1#2\par{\expandafter\xdef\csname @ref@#1\endcsname{#2}}



\def\data{\number\day.\number\month.\number\year}
\def\datasotto{\@datatrue
\footline={\hfil{\rm \data}\hfil}}

\def\titoli#1{\if@data\relax\else\footline={\hfil}\fi
         \xdef\prima@riga{#1}\voffset+20pt
        \headline={\ifnum\pageno=1
             {\hfil}\else\hfil{\sl \prima@riga}\hfil\folio\fi}}

\def\duetitoli#1#2{\if@data\relax\else\footline={\hfil}\fi
         \voffset=+20pt
    \headline={\ifnum\pageno=1
             {\hfil}\else{\ifodd\pageno\hfil{\sl #2}\hfil\folio
\else\folio\hfil{\sl #1}\hfil\fi}  \fi} }

\def\la@sezionecorrente{0}


\def\const@label#1{\global\advance\const@num by 1\xdef            
   \la@costante{\the\const@num}\expandafter
   \xdef\csname @const@#1\endcsname{\la@costante}}

\def\cconlabel#1{\se@indefinito{@const@#1}
\const@label{#1}\fi}

\def\constnum#1{\se@indefinito{@const@#1}
\const@label{#1}\fi\csname @const@#1\endcsname}

\def\ccon#1{C_{\constnum{#1}}}

\catcode`@=12


\def\abstract{
\vskip48pt plus 60pt
\noindent {\bf Abstract.}\quad}

\def\theorem#1#2{\par\vskip4pt
\noindent {\bf Theorem \autolemma{#1}.}{\sl \ #2}
 \par\vskip10pt}

\def\lemma#1#2{\par\vskip4pt
\noindent {\bf Lemma \autolemma{#1}.}{\sl \ #2}
 \par\vskip4pt}

\def\proof{\par\noindent{\bf Proof.}\ }

\def\corollary#1#2{\par\vskip4pt
\noindent {\bf Corollary \autolemma{#1}.}{\sl \ #2}
 \par\vskip10pt}


\def\norma#1{\left\Vert#1\right\Vert}

\def\frac#1#2{{#1\over #2}}

\def\interno{\vbox{\hbox{\vbox to .3 truecm{\vfill\hbox to .2 truecm
{\hfill\hfill}\vfill}\vrule}\hrule}\hskip 2pt}

\def\quadratino{
\hfill\vbox{\hrule\hbox{\vrule\vbox to 7 pt {\vfill\hbox to 7 pt
{\hfill\hfill}\vfill}\vrule}\hrule}\par}

\def\ponesotto#1\su#2{\mathrel{\mathop{\kern0pt #1}\limits_{#2}}}

\def\tdot#1{\hskip2pt\ddot{\null}\hskip2.5pt \dot{\null}\kern -5pt {#1}}

\def\charslash#1{\setbox2=\hbox{$#1$}
     \dimen2=\wd2
     \setbox1=\hbox{/}\dimen1=\wd1
     \ifdim\dimen2>\dimen1
     \rlap{\hbox to \dimen2{\hfil /\hfil}}
     #1
     \else
     \rlap{\hbox to \dimen1{\hfil$#1$\hfil}}
     /
     \hfil\fi}

\def\Re{{\rm \kern 0.4ex I \kern -0.4 ex R}}

\def\poisson#1#2{\left\{#1 ,#2\right\} }

\def\ra{{\bf Z}}

\def\Cm{{\bf C}}

\def\imma{{\rm Im}\hskip2pt}

\def\pmb#1{\setbox0=\hbox{#1}\ignorespaces
    \hbox{\kern-.02em\copy0\kern-\wd0\ignorespaces
    \kern.05em\copy0\kern-\wd0\ignorespaces
    \kern-.02em\raise.02em\box0 }}

\def\A{{\cal A}}

\def\G{{\cal G}}
\def\H{{\cal H}}

\def\U{{\cal U}}
\def\V{{\cal V}}

\def\uno{{\kern+.3em {\rm 1} \kern -.22em {\rm l}}}


\newskip\ttglue


\riferimentifuturi
\autobibliografia

\numerazionedoppia
\def\Ne{{\rm \kern 0.4ex I \kern -0.4 ex N}}
\def\Ce{{\rm \kern 0.4ex I \kern -0.4 ex C}}
\def\ha{Ha\-mil\-to\-nian}
\def\la{\langle}
\def\ra{\rangle}
\def\cl{\A}
\def\mod#1{\left|#1\right|}
\def\op#1{Op^w\left(#1\right)}
\def\h{\H}
\def\De{\Delta_{\h}}
\def\moyal#1#2{\poisson{#1}{#2}_M}
\def\indi#1{t-\tau_1,\tau_1,\tau_2,...,\tau_{#1}}
\def\st{\sigma e^{-\alpha |t|}}
\def\rt{\rho e^{-\alpha |t|}}
\def\ds{\displaystyle}

\def\Om{\Omega}

\def\Ha{Ha\-mil\-to\-nian}
\centerline{\bf   LONG TIME SEMICLASSICAL APPROXIMATION OF QUANTUM FLOWS:}
\centerline{\bf A PROOF OF THE EHRENFEST TIME}
\vskip 24pt
\centerline{Dario BAMBUSI\footnote{$^1$}
{\sm Dipartimento di Matematica, Universit\`a di Milano,
Italy. (bambusi@mat.unimi.it)}, Sandro GRAFFI\footnote{$^2$}
{\sm Dipartimento di Matematica, Universit\`a di Bologna,
Italy. (graffi@dm.unibo.it)}, Thierry PAUL\footnote{$^3$}
{\sm CEREMADE, Universit\'e de Paris-Dauphine,
France. (paulth@ceremade.dauphine.fr)}}
\abstract{\it Let $ \h$ be a holomorphic \Ha\ of quadratic growth on
$ \Re^{2n}$, $b$ a holomorphic exponentially localized observable,
$H$, $B$ the corresponding operators on $L^2(\Re^n)$ generated by Weyl
quantization, and $U(t)=\exp{iHt/\hbar}$. It is proved that the $L^2$
norm of the difference between the Heisenberg observable
$B_t=U(t)BU(-t)$ and its semiclassical approximation of order
${N-1}$ is majorized by $K N^{(6n+1)N}(-\hbar{\rm log}\hbar)^N$
for $t\in [0,T_N(\hbar)]$ where $
\ds T_N(\hbar)=-{2{\rm log}\hbar\over {N-1}}$. Choosing a suitable
$N(\hbar)$ the error is majorized 
 by $C\hbar^{\log|\log\hbar|}$, $0\leq t\leq |\log\hbar|/\log|\log\hbar|$.
(Here $K,C$ are constants independent of $N,\hbar$).}

\autosez{p}Introduction and statement of the results

Denote $\Omega:=\Re^{2n}$ with coordinates  $(x,\xi)$. Let 
 $\h(x,\xi)\in C^{\infty}(\Omega;\Re)$, and 
$b_t(x,\xi):=b\circ \phi^{\h}_t\equiv b(\phi^{\h}_t(x,\xi))$ be the time
evolution of any bounded observable
$b(x,\xi)\in C^{\infty}(\Omega;\Re)$ under the the flow
 $\phi^{\h}_t:\Om\leftrightarrow\Om$ generated by the \ha\ $\h$. Denote 
$H:=Op^W(\h)$ and $B=Op^W(b)$ the self-adjoint operators in $L^2(\Re^n)$
representing the (Weyl) quantization of the symbols $\h,b$ and let $\ds
B_t:=e^{iHt/\hbar}Be^{-iHt/\hbar}$ be the Heisenberg observable, i.e. the 
quantum evolution of the observable
$B$ under the unitary group generated by
$H$. \par 
The question of estimating how long the classical and quantum
evolutions stay "close" one another or, better, how long the evolution of
the quantum observables is determined by the corresponding classical one up to a
prescribed error vanishing with $\hbar$ is one of the oldest problems of
semiclassical analysis.  According to a well known conjecture going back to
Chirikov and Zaslavski [Ch,Za], this approximation can be valid on a time interval
of maximum duration
$T\equiv T(\hbar)$  of order
$-{\rm log}\hbar$, called the Ehrenfest time, if the error is required to vanish
faster than any power of
$\hbar$.

The origin of this conjecture, formally verified in some instances[Za] can be
understood   in the correspondence between symbols $b(x,\xi)$ (classical
observables) and operators in Hilbert space $B$ (quantum observables) provided by
the Weyl quantization procedure:
$$
(Bu)(x)={1\over (2\pi\hbar^n)}
\int_{\Re^{2n}}b\left({{x+y}\over 2},\xi\right)e^{i\langle
(x-y),\xi\rangle/\hbar}\,u(y)\,dyd\xi,
\quad u\in{\cal S}(\Re^n) \autoeqno{Weyl}
$$
In this framework the problem can be formulated as follows: 
$B_t$ solves the Heisenberg equation of motion 
$$
\dot{B}_t={i\over\hbar}[H,B_t] \autoeqno{Heisenberg}
$$
If $B_t$ admits a symbol, denoted $b_t(x,\xi;\hbar)$, by
\eqref{Heisenberg} it fulfills the  equation
$$
\dot{b}_t=\{\h,b_t\}_M \autoeqno{Moyal}
$$
with the initial condition $b_0(x,\xi;\hbar)=b(x,\xi)$. Here  $\{f,g\}_M(x,\xi)$
is the Moyal bracket of the two observables
$f,g\in C^{\infty}(\Re^{2n})$
$$
\{f,g\}_M(x,\xi):=f\#g-g\#f \autoeqno{Moyaldef}
$$
where $f\#g$,  the symbol of the operator product $FG$, is expressed by the  
composition of the symbols $f$ and $g$:
$$
(f\#g)(x,\xi)={1\over (2\pi\hbar)^{n/2}}\int_{\Re^{4n}}e^{-i\la
r,\rho\ra/\hbar\,+i\la w,\tau\ra/\hbar}\,f(x+w,\rho+\xi)g(x+r,\tau+\xi)\,d\rho
d\tau\,dr dw
\autoeqno{composizione}
$$
$\{f,g\}_M$  admits the following formal expansion in
powers of
$\hbar$ [Fo,Ro,Vo]:
$$
\{f,g\}_M(x,\xi)\sim\{f,g\}+{1\over2^j}\sum_{|\alpha+\beta|=j\geq
1}(-1)^{|\beta|}\hbar^j
\left(\partial^{\alpha}_{\xi}gD^{\beta}_xg\right)\cdot
\left(\partial^{\beta}_{\xi}gD^{\alpha}_xf\right) \autoeqno{Moyalexp}
$$
($\alpha=(\alpha_1,\ldots,\alpha_n)$ is a multi-index, and 
$|\alpha|:=\alpha_1+\ldots+\alpha_n$; analogous definitions for
$\beta$, and $\ds D_x:=-i\hbar\partial_x$). By  \eqref{Moyalexp} 
the differential equation
\eqref{Moyal} can be recursively solved in the space of the formal power
series in $\hbar$ (for details see [Ro], Chapt.IV.10). The result, known as the
semiclassical Egorov theorem, is the formal semiclassical
expansion of the symbol $b_t$:
$$
b_t(x,\xi;\hbar)\sim (b\circ
\phi_t^a)(x,\xi)+\sum_{j=2}^{\infty}b_j(x,\xi;t)\hbar^j \autoeqno{Eg1}
$$
Here the term of order zero in $\hbar$, by definition the 
{\it principal symbol} of $B_t$, is
just the evolution of the observable $b$ along the \Ha\ flow generated by $\h$,
i.e. the solution  of the Liouville equation $\dot{b}_t=\{\h,b_t\}$, and
$$
b_j(x,\xi;t)=-i\int_0^t\sum_{|\alpha+\beta|+l=j+1 \atop 0\leq l\leq j-1}
(1-(-1)^{|\alpha+\beta|})\Gamma(\alpha,\beta)
\left(\partial^{\alpha}_{\xi}\h D^{\beta}_xb_l\right)\circ
\phi^{\h}_{t-\tau}(x,\xi)\,d\tau \autoeqno{Eg2}
$$
 The higher order
terms $b_j(x,\xi;t)$ are thus completely determined by the classical
evolution but have a polynomial dependence on the derivatives of the flow
$\phi^{\h}_t(x,\xi)$ with respect to the inital conditions
$(x,\xi)$ up to order $j-1$. If, as it  happens in general, there are
initial conditions $(x,\xi)$ generating a flow with positive Lyapunov exponents,
the difference between the symbol $b_t(x,\xi;\hbar)$ of $B_t$ and any
prescribed approximation $\ds (b\circ
\phi_t^{\h})(x,\xi)+\sum_{j=2}^{N}b_j(x,\xi;t)\hbar^j$ is expected to increase
exponentially in time: hence it can vanish as $\hbar\to 0$ only for a time
interval not exceeding
$-{\rm log}\hbar$. Put in a different way: the  non-local nature
of quantum mechanics, embodied in the symbol expansion
\eqref{Eg1}, \eqref{Eg2}, can be dominated by its local
approximation, the principal symbol $b\circ \phi_t^{\h}(x,\xi)$, only if the
 the remainder is small. This can be obtained only within the above time
span. \par In this paper we work out, in the analytic case, the
estimates implying the validity of the above "Ehrenfest time" for a class of flows
somewhat restricted but in a sense natural as discussed below. More
precisely, for any fixed
$\sigma >0$ set  $|z|:=\sup|z_k|$ and 
$\ds 
\G_{\sigma}:=\left\{z\in \Cm^{2n} \ :\ |\imma z|<\sigma\right\}$. 
 The \Ha\  $\h(x,\xi)\equiv
\h(z)$, $z:=(x,\xi)$ is required to fulfill the following
properties:\par\noindent
\item {(A1)} There exists $\nu>0$ such that $\h$ is real-holomorphic on
$\G_{\nu}$.
\item{(A2)} Let $Jd\h$ be the symplectic gradient of $\h$. Then there 
are  $A_1>0, A_2>0, \alpha>0$ such that $\left|Jd\h(z+iy)\right|\leq
A_1+A_2|z|$  
$\forall z, y\in\Re^{2n}$, $\left|y\right|\leq\sigma$. Moreover 
$|Jd^2\h(z)|\leq \alpha$ on $\G_\sigma$.
\item{(A3)} Denote $(\hat\h)(k)$ the Fourier transform of $\h(z)$. Then there
are  $\rho>0, \sigma>0$ such that
$\hat\h(k_1+ik_2)$  is holomorphic on $\G_{\rho}\setminus (0,0)$; moreover  
$k^3\hat\h(k)$ is holomorphic on
$\G_{\rho}$ and 
$$
|k_1|^3|\hat \h(k_1+ik_2)|\leq C e^{-\sigma|k_1|} \quad {\rm for}\ 
|k_2|\leq \rho
$$ 
\par\noindent
{\bf Remarks}.
\item 1 Under the above assumptions $H=Op^W(\h)$ defined by \eqref{Weyl} is
essentially self-adjoint in $L^2(\Re^n)$. By a standard abuse of notation
we denote $H$ also its self-adjoint closure.
\item 2 Within the analyticity and decay
assumptions (A1)-(A3),  (A2) is  the quadratic
growth  condition  ensuring the existence of the  Fourier integral operator
representing the pro\-pa\-gator
$\ds
\exp{iHt\over
\hbar}$ [Cha] and thus the existence of the symbol of $B_t$ [Ro].
\item 3 In the phase variables $z=(x,\xi)$ Assumption (A3) means that there are
$\sigma,\rho>0$ such that 
$$
\sup_{z+iy\in
\G_{\sigma}}\mod{\partial^{|\alpha|+|\beta|=3}_z\h(z+iy)}e^{\rho|z|}<+\infty
.\autoeqno{n.13}
$$
\vskip 0.1cm\noindent
To state the main result of the paper we need some further
notation. For $b$ as above  set:
$$
\De b:=\frac1{\hbar^2}\left[\poisson b{\h}-\moyal{b}{\h}\right] \ .\autoeqno{De}
$$
and define recursively the two sequences $r_k^t, b^t_k: k\geq 1$ in the following
way:
$$
r_1^{t-\tau_1}:=\De (b\circ \phi^{t-\tau_1}),\quad
r_{k+1}^{\indi{k}}:=\De\left[r_k^{\indi{k-1}}\circ\phi^{\tau_k}\right] 
\autoeqno{b_j}
$$
$$
b_k^t:=\int_0^td\tau_1\int_0^{t-\tau_1}d\tau_2\int_0^{t-\tau_2}d\tau_3...
\int_0^{t-\tau_{k-1}}d\tau_kr_k^{\indi{k-1}}\circ \phi^{\tau_k};
b_0:=b\circ\phi^t
$$
Moreover, let $b:\G_{\sigma}\to \Cm$ be holomorphic. Set:
$$
\mod b_{\sigma,\rho}:=\sup_{x+iy\in
\G_{\sigma}}\mod{b(x+iy)}e^{\rho|x|}\ .\autoeqno{n.1}
$$
Denote $\cl_{\sigma,\rho}$  the set of all functions $f$ 
holomorphic on $\G_{\sigma,\rho}$ such that $\mod f_{\sigma,\rho}<+\infty$.
Then:
\theorem{1.1}{Let there exist $\sigma >0, \rho >0$ and
$0<\overline{B}<+\infty$ such that
$\mod b_{\sigma,\rho} <\overline{B}$. Then: 
\item{(1)} The operators $B^j_t:=Op^W(b_j^t)$ are continuous in
$L^2$ and the Heisenberg operator $B_t=U(t)BU(-t)$ admits the expansion
$$
B_t=\sum_{j=0}^NB_j^t\hbar^{2j}+\hbar^{2(N+1)}S_N^t\ ,
$$
where
$$
S_N^t:=\int_0^td\tau_1\int_0^{t-\tau_1}d\tau_2\int_0^{t-\tau_2}d\tau_3...
\int_0^{t-\tau_{k-1}}d\tau_k U(\tau_k)\op{r_k^{\indi{k-1}}}U(-\tau_k)
$$
\item{(2)} There are positive constants $E, F$ independent of $j,N$
and $\hbar$ such that
for all $j\geq 1$, $N\geq 2$, $t\geq 0$ the following estimates hold:
$$
\norma{B^t_j}_{L^2\to L^2}\leq
\left[eEe^{7\alpha t}j^{6n+3}\right]^j {\overline{B}F\over 
j!}e^{(4n+2)\alpha t}
\left[\exp\left( \alpha\frac{j(j-1)}2t\right)\right]^{6n+3} \autoeqno{stimaj}
$$
$$
\norma{S_N^t}_{L^2\to L^2}\leq
\left[eEe^{7\alpha t}N^{6n+3}t\right]^N 
\frac{\overline{B}F}{N!}e^{(4n+2)\alpha t}
\left[\exp\left( \alpha\frac{N(N-1)}2t\right)\right]^{6n+3} \autoeqno{stimaresto}
$$} 
\noindent {\bf Remark.} The holomorphy assumptions are needed to control
the remainder to order $\hbar^N$ for all $N$. If we limit ourselves to
$N=1$ more general classes of
\ha s and of observables can be considered. More precisely, let for
instance $\h(x,\xi)$ be a polynomial of order $2p$ such that the subgraph
$\Sigma_E:=\{(x,\xi)\in\Re^{2n}|\h(x,\xi)\leq E\}$ is compact for some
$E$, and let $b(x,\xi)\in C_0^{\infty}(\Sigma_{{E}})$. 
Then (proof in the next section)  there are $\Gamma>0$ and $\Delta>0$ such that:
$$
\norma{B^t-B^t_0}_{L^2\to L^2}=\norma{B^t-Op^W(b\circ\phi^t)}_{L^2\to
L^2}\leq \Gamma\hbar^2 t e^{\Delta t}\;. \autoeqno{stima1ord}
$$
\vskip 0.4cm\noindent
The symbols
$b_j^t$ and hence the operators
$B_j^t$ are completely
determined by the classical flow $\phi^t$ via \eqref{b_j}. The quantum
evolution will then stay close to the (semi) classical one as long as
the error $S_N^t$ stays small.  The estimate
\eqref{stimaresto} yields indeed, through a straightforward computation:
\corollary{1.12}{Let $\ds T_N(\hbar):=-{2{\rm log}\hbar\over \alpha(N-1)}$, 
  $\ds B^N_t:=\sum_{j=O}^{N-1}B_j^t\hbar^j$. Then, for $0\leq t\leq
T_N(\hbar)$: 
$$
\norma{B_t-B^N_t}\leq  \left(\frac{2e^2E}{\alpha }\right)^N
\,N^{(6n+1)N}BF \hbar^{2-15/\alpha-(8n+4)/\alpha N}
(-\hbar{\rm log}\hbar)^N \autoeqno{stimaN}
$$
}
\noindent
{\bf Remarks.}
\item{1} If the Lyapunov numbers are zero for any initial datum
$(x,\xi)$, then we can take $\alpha=0$ in formula \eqref{stimaresto}, 
and by Assertion 2 of Theorem 1.2 one has
$$
\norma{B_t-B^{N(\hbar)}_t} =O(\hbar^N) \qquad 0\leq t\leq
\tilde{T}_N(\hbar)
$$
where $\ds 
\tilde{T}_N(\hbar):=  e^{-1} N^{-6n-1}h^{-1}$. 
\item{2}
Estimates valid for a time interval of duration $-C_N{\rm log}\hbar$  for
Hamiltonians admitting  polynomial growth of any order (but without control of
the constant $C_N$),  have been obtained by Combescure and Robert  [Co-Ro1] in a
weaker sense, i.e.  comparing  classical and quantum evolutions along coherent
states (according to ideas introduced in [He], [BZ] 
and developed in [Ha], [BIZ], [Co-Ro2]).
\item {3} The symbol expansion generated by Assertion (1) of Theorem
\lemmaref{1.1}, namely
$$
b_t(x,\xi;\hbar)=\sum_{j=0}^N b^t_j\hbar^{2j}+O(\hbar^{2N+1})
$$
differs from \eqref{Eg1} in all terms with $j>0$. 
This difference makes the present expansion a non formal one, so that its 
remainder can be estimated.
\item{4} Finally, let $T(\hbar)\in C([0,1];\Re_+)$ be an
increasing function such that $\ds \lim_{\hbar\to 0}\frac{T(\hbar)}{-{\rm
log}\hbar}=0$, and let 
$\ds N(\hbar):=\left[-\frac{{\rm log}\hbar}{T(\hbar)}\right]$. Then
clearly $\norma{B_t-B^{N(\hbar)}_t} =O(\hbar^{\infty}) $, $0\leq t\leq
T(\hbar)$.
\vskip 0.1cm\noindent
To put this result into a more quantitative version, define the function
sequence $\{\log^{[k]}(x)\}: k\in\Ne$ by
$\log^{[1]}(x):=\log(x)$, $\log^{[k]}(x):=\log(\log^{[k-1]}(x))$.

\corollary{hinfinito}{For any integer $k\geq 1$ define
$N_k(\hbar):=[\log^{[k]}(|\log(\hbar)|)]$. Then there exist positive
constants $C$, $\overline{\hbar}$ such that, for 
$0<\hbar\leq\overline{\hbar}$ one has
$$
\norma{B_t-B^{N_k(\hbar)}_t}
\leq C\; \hbar^{\log^{[k]}(|\log(\hbar)|)}
$$
for
$$
0\leq t\leq \frac{|\log(\hbar)|}{\log^{[k]}(|\log(\hbar)|)}
$$
}
\noindent{\bf Acknowledgments.} We thank A.Martinez and D.Robert for reading the
paper and several useful remarks. We acknowledge the support of
CEREMADE that made possible the collaboration leading to this work.

\autosez{p1}Proofs

Let $b$ be a Weyl symbol of class $\Sigma_0^1$, and $\h$ an admissible
semiclassical symbol  (For these notions, see [Ro], Chapter 2;
particular examples are all bounded observables
$b(x,\xi)\in C^{\infty}(\Re^{2n})$ and the Hamiltonians $\h\in
C^{\infty}(\Re^{2n})$ of polynomial growth at infinity).  Denote
$\phi^t$ the flow generated by
$Jd\h$, $J$ the unit $2n\times 2n$ symplectic matrix; let $H=\op{\h}$ be
essentially 
 self-adjoint in $L^2(\Re^n)$ and denote also  
$U(t):=\exp(itH/\hbar)$,
$B:=\op b$,
$B_t:=U(t)BU(-t)$, and
$$
\De b:=\frac1{\hbar^2}
\left[\poisson b{\h}-\moyal{b}{\h}\right] \ .\autoeqno{De.1}
$$
Our semiclassical expansion is generated by the following simple remark:

\lemma{1.2}{The following formula holds
$$
B_t:=\op{b\circ\phi^t}+\hbar^2\int_0^td\tau
U(\tau)\op{r_1^{t-\tau}}U(-\tau)\ ,
$$
where $r_1^s:=\De\left(b\circ \phi^s \right)$}

\proof Denote $\beta_t:=b\circ \phi^t$. Then:
$$
\eqalign{
\frac d{dt}\left[\op{\beta_t}\right]=\op{\frac d{dt}\beta_t}=
\op{\poisson{\beta_t}{\h}}
\cr
=\op{\poisson{\beta_t}{\h}-\poisson{\beta_t}{\h}_M}+\frac
i{\hbar}\left[\op{\beta_t}, \op{\h}\right]
\cr
=\frac
i{\hbar}\left[\op{\beta_t}, \op{\h}\right]+\hbar^2\op{r_1^t}\ .
}$$
It follows
$$
\frac d{dt}\left[\op{\beta_t}-B_t\right]=\frac
i{\hbar}\left[\op{\beta_t}-B_t, \op{\h}\right]+\hbar^2\op{r_1^t}\ ,
$$
and by the variation of parameters  formula  
$$
\op{\beta_t}-B_t=\hbar^2\int_0^t U(t-s)\op{r_1^s}U(-(t-s))ds\ .
\autoeqno{diff1}
$$
The assertion is now proved performing the change of variable  
$\tau=t-s$  in the integral.\quadratino 
\vskip 0.1cm\noindent
{\bf Proof of formula \eqref{stima1ord}}. Since $\h$ is a polynomial
of degree $2p$ 
$$
r_1^t=\De (b\circ\phi^t)=\frac1{\hbar^2}\left[\poisson
{b\circ\phi^t}{\h}-\moyal{b\circ\phi^t}{\h}\right]
=\sum_{{|k|=1\atop k=(k_1,...,k_{2n})}}^{2p}c_k\hbar^{k}
\frac{\partial^{|k|}b\circ\phi^t}{\partial
z^k}
$$
where $c_k(x,\xi)$ is a polynomial of degree $2p-|k|$.  Now the
smooth functions $\ds \theta_k(x,\xi):=c_k(x,\xi)\frac{\partial^{|k|}
b\circ\phi^t}{\partial z^k}$ have
compact support in
$\Re^{2n}$ and hence define bounded operators in $L^2$ upon Weyl
quantization. Denote $\lambda(x,\xi)$ the Lyapunov number of the
trajectory $\phi^t$ with any initial datum
$(x,\xi)\in\Sigma_{{E}}$. Since $\phi^t(x,\xi)$ is bounded
$\forall\,t\in\Re$ we have (see e.g.[Ce], 3.12) $\ds \delta:={\rm
sup}_{\Sigma_{{E}}}\lambda(x,\xi)<+\infty$. Hence there are
$\gamma_k>0$ such that $\ds {\rm
sup}_{\Sigma_{{E}}}\left|\frac{\partial^{|k|}\phi^t}{\partial
z^k}\right|\leq
\gamma_ke^{\delta t}$. Since  $\ds \frac{\partial^{|k|}b\circ\phi^t}{\partial
z^k}$ is a polynomial of degree $|k|$ in the variables 
$\ds \frac{\partial^{|s|}\phi^t}{\partial
z^s},\, s=1,\ldots,|k|$  with coefficients depending 
on $\ds \frac{\partial^{|s|}b}{\partial
z^s},\, s=1,\ldots,|k|$, for any fixed $q\in\Ne$ depending only on $n$ 
there are $\Gamma_k(n)>0$ such  $\ds {\rm
sup}_{\Re^n}\left|\frac{\partial^{|l|}\theta_k}{\partial
z^l}\right|\leq \Gamma_ke^{|k|q\delta t}, |l|\leq q$. 
Hence by the Calderon-Vaillancourt theorem  there exists $q>0$ such that 
$ \ds \|Op^W(\theta_k)\|_{L^2\to L^2}\leq
\Gamma_ke^{|k|q\delta t}$. Inserting this estimate in \eqref{diff1} we get
\eqref{stima1ord} with 
$\Delta=2pq\delta, \Gamma={\rm Max}_k \Gamma_k$ .\quadratino
\vskip 0.2cm\noindent
  Recall now the definition of the sequences
$r_k^{\indi{k-1}}$ ($k\geq2$) and 
  $\left\{b_k^t\right\}_{k\geq 0}$,  $b_0:=b$:
$$
\eqalign{
r_{k+1}^{\indi{k}}&:=\De\left[r_k^{\indi{k-1}}\circ\phi^{\tau_i}\right]
}
$$
$$
b_k^t:=\int_0^td\tau_1\int_0^{t-\tau_1}d\tau_2\int_0^{t-\tau_2}d\tau_3...
\int_0^{t-\tau_{k-1}}d\tau_kr_k^{\indi{k-1}}\circ \phi^{\tau_k}\;\; k\geq 1
$$

\lemma{1.3}{Let $B^t_j=Op^W(b^t_j)$. Then:
$$
\eqalign{
B_t=\sum_{j=0}^NB_j^t\hbar^{2j}+\hbar^{2(N+1)}S_N\ ,
}
$$
where
$$
S_N:=\int_0^td\tau_1\int_0^{t-\tau_1}d\tau_2\int_0^{t-\tau_2}d\tau_3...
\int_0^{t-\tau_{k-1}}d\tau_k U(\tau_k)\op{r_k^{\indi{k-1}}}U(-\tau_k)
$$
}
\proof Just iterate the proof of lemma \lemmaref{1.2}\quadratino 
\vskip 0.1cm\noindent
Let $b:\G_{\sigma}\to \Cm$ be an analytic
function; recall the definitions
$$
\mod b_{\sigma,\rho}:=\sup_{x+iy\in
\G_{\sigma}}\mod{b(x+iy)}e^{\rho|x|}\ .\autoeqno{n.111}
$$
and $\cl_{\sigma,\rho}:=\{b\;{\rm holomorphic}\;{\rm in}\;
\G_{\sigma}: \mod b_{\sigma,\rho} <+\infty\}$. We 
will estimate the sequence $r_k$ in the above norm. Clearly we have
to estimate the norm of $b\circ \phi^t$ and of $\De b$. We first prove
the following 

\lemma{phi1}{There exists a positive $\sigma$ such that
$\phi^t$ extends to a complex analytic function 
$$
\phi^t:\G_{\sigma e^{-\alpha t}}\to \G_{\sigma}
$$
}
\proof Denote $f:=Jd\h$, and consider, on $\G_{\sigma}$, 
the system of equations
$$
\dot z=f(z) \ .\autoeqno{1.2}
$$
Writing $z=x+iy$ and $f=f_1+if_2$,
one has $\dot y=f_2(x+iy)$. Since $f_2=0$ on the real axis, by
assumption A2 one has $|f_2(x+iy)|\leq \alpha |y|$. It follows that
the inequalities
$$
|y|\dot{\null}\leq\alpha|y|\quad\Longrightarrow |y(t)|\leq
|y_0|e^{\alpha |t|} \autoeqno{1.1}
$$
hold. So one has $\phi^t(\G_{\st})\subset \G_\sigma$.

Fix $\tilde t$. Given $\bar z\in \G_{\sigma e^{-\alpha|\tilde t|}}$ we
prove that $\phi^{\tilde t}$ is analytic at $\bar z$. By the Cauchy-
Kowaleskaya theorem (see e.g.[Pe]) there exists a neighbourhood $ \U$ of $\bar
z$  and a time $\bar t$ such that,
for any $|\tau|<\bar t$, $\phi^\tau$ is analytic on $\U$. Assume that
$\bar t$ is the supremum of such times (so that $\phi^{\bar t}$ is
not analytic in $\U$). Assume by contradiction $\bar t<\tilde t$. By
\eqref{1.1} the limit $\lim_{\tau\to\bar t}\phi^\tau(z)$ exists on
$\U$. Denote $w :=\lim_{\tau\to\bar t}\phi^\tau(\bar z)$. Again by
the Cauchy-Kowakeskaya thoerem there exists a neighbourhood $\V$ of $w$ and
a $t_1>0$ such that $\phi^{\tau}$ is analytic on $\V$ for
$\left|\tau\right|<t_1$. Assume that $\U$ is so small that for fixed
$\epsilon$ small enough one has $\phi^{\bar t-\epsilon}(\U)\subset
\V$, then one has 
$$
\phi^{\bar t+\epsilon}(\U)=\phi^{2\epsilon}\left(\phi^{\bar
t-\epsilon}(\U) \right)\ ,
$$    
which is analytic since it is the composition of two analytic
functions, against the assumption that $\bar t$ is the last time of
analyticity. \quadratino 

\lemma{phi}{Let $b\in \cl_{\sigma,\rho}$, then, for any $t$, and for
$\sigma$ small enough, one has $b\circ\phi^t\in\cl_{\st,\rt}$, and
$$
\mod{b\circ\phi^t}_{\st,\rt}\leq \mod b_{\sigma,\rho}\ .
$$
}

\proof By the above lemma $b\circ\phi^t$ has the required analyticity
properties. Denote $\rho_t:=\rt$, $\sigma_t:=\st$,
$\varphi_1+i\varphi_2=\phi^t(x+iy)$, then one has
$$
\eqalign{
\mod{b\circ\phi^t}_{\sigma_t,\rho_t}=\sup_{x+iy\in
\G_{\sigma_t}}\left|b\left( \phi^t(x+iy)\right)e^{\rho_t|x|} \right|
\cr
\leq
\sup_{\varphi_1+i\varphi_2\in\G_{\sigma_te^{\alpha t}}}\left|
b(\varphi_1+i\varphi_2 )e^{\rho_t|Re(\phi^{-t}(\varphi_1+i\varphi_2
))|}\right|\ ;
}
$$
using the equation of motion and A2, one has
$$
|Re(\phi^{-t}(\varphi_1+i\varphi_2
))|<|\varphi_1|e^{\alpha|t|}\ ,
$$
 which implies the assertion. \quadratino

We will estimate the norm of $\De$ using the Fourier transform. For
this reason the following lemma is useful

\lemma{hat}{One has
$$
\mod{\hat
b}_{\rho-\delta,\sigma}\leq\left(\frac2\pi\right)^{n}\frac1{\delta^{2n
}}\mod{b}_{\sigma,\rho}\ .\autoeqno{h.3}
$$
}

\proof Fix $k_1=\kappa e_1$ where $e_1$ is the unit vector of the
first axis and $\kappa$ a positive number; fix also
$k_2$ with $|k_2|<\rho-\delta$. One has
$$
\eqalign{
(2\pi)^{n}\left|\hat b(k_1+ik_2)\right|= \left|\int _{\Re^{2n}}
b(x)e^{i(k_1+ik_2)x} dx \right|
\cr
= \left|\int _{\Re^{2n}}
b(x+ie_1\sigma)e^{i(k_1+ik_2)(x+ie_1\sigma)} dx \right|
\leq
\int_{\Re^{2n}}|b|_{\sigma,\rho}e^{-\rho|x|}e^{-\kappa\sigma}e^{|k_2||x|}
\cr
\leq
\left|b\right|_{\sigma,\rho}e^{-\kappa\sigma}\int_{\Re^{2n}}e^{
-\delta|x|}dx
=\left(\frac2\delta\right)^{2n}e^{-\kappa\sigma}|b|_{\sigma,\rho}\ ,
}
$$
which by definition of $\ds \mod{b}_{\sigma,\rho}$ is the thesis in the
particular case just considered. The general case can be dealt with in a similar
way.
\quadratino
\vskip 0.1cm\noindent
\lemma{delta}{Let $b\in\cl_{\sigma,\rho}$ with $\sigma\leq\nu$ small
enough. Then there exists a positive constant $A$ such that,
$\forall d<\sigma$, $\delta<\rho$:
$$
|\De b|_{\sigma-d,\rho-\delta}\leq \frac
A{\delta^{2n}d^{4n+3}}|b|_{\sigma,\rho} \ .
$$
} 
\proof  To obtain the estimate via the Fourier transform we first recall that 
$$
\moyal b{\h}^{\wedge}(k)=\frac2\hbar \int_{\Re^{2n}} \hat
b(k-s)\hat\h(s)
\sin{\frac{(k-s)\wedge s}
{\hbar/2}}ds
$$
where $(k_p,k_q)\wedge(s_p,s_q):=k_p\cdot s_q-k_q\cdot s_p$, whence
$$
\widehat{\De b}(k) =\frac2\hbar \int_{\Re^{2n}} \hat
b(k-s)\hat\h(s)
\left(\sin{\frac{(k-s)\wedge s}
{\hbar/2}}-\frac{(k-s)\wedge s}
{\hbar/2}\right)ds
$$
Since $|\sin z-z|\leq\ccon 1 |z^3|$  for all
$z\in\G_\sigma$, one has, for $|\imma k|<\rho-\delta$,
$$
\eqalign{
\left|\widehat{\De b}(k)\right|\leq \int_{\Re^{2n}}\left|\hat
b(k-s)\right|
\left|\hat \h(s)\right|\ccon 1|k-s|^3|s|^3ds
\cr
\leq \ccon 2
\frac{|b|_{\sigma,\rho}}{\delta^{2n}}\int_{\Re^{2n}}|k_1-s|^3e^{-\sigma|k_1-s|}
e^{-\sigma|s|}ds\ ,
}\autoeqno{a.w}
$$ 
where  use has been made of \eqref{h.3} and Assumption A2. 
Now $\left| |k_1-s|+|s|\right|\geq |k_1|$
and $\left| |k_1-s|+|s|\right|\geq |k_1-s|$. Hence \eqref{a.w} does not exceed 
$$
\eqalign{
\frac{|b|_{\sigma,\rho}}{\delta^{2n}}\ccon a e^{-(\sigma-d)|k_1|}
\int_{\Re^{2n}}|k_1-s|^3 e^{-d|k_1-s|}ds
\cr
=\frac{|b|_{\sigma,\rho}}{\delta^{2n}d^{2n+3}}\ccon a
e^{-(\sigma-d)|k_1|}
\int_{\Re^{2n}}|s|^3e^{-|s|}ds\ ,
}
$$
which gives 
$$
\left|\widehat{\De b}\right|_{\rho-\delta,\sigma-d}\leq \frac{\ccon
b}{d^{2n+3} \delta^{2n}} |b|_{\sigma,\rho}\ .
$$
Using again \eqref{h.3} to antitransform $\widehat{\De b} $ 
the assertion is proved.\quadratino
\vskip 0.1cm\noindent

\lemma{r.k}{Assume $|b|_{\sigma,\rho}\leq \overline{B}$ for some
positive $\overline{B},\sigma,\rho$. Then, for $k\geq 1$ and
$0<\tau_k<t$, one has
$$
\left|r_k^{\indi{k-1}}\right|_{(\sigma-k\delta)e_{k},(\rho-kd)e_{k}}\leq
\Gamma_k
$$
Here the sequence $e_k$ is defined by
$$
e_1:=e^{-\alpha t}\ ,\quad e_k:=e_1\exp\left(-\alpha t(k-1)\right)\
,\quad k\geq2 \autoeqno{e.k}
$$
and the sequence $\Gamma_k$  by
$$
\eqalign{
\Gamma_1&:=\frac{A\overline{B}}{\delta^{2n}d^{4n+3}}\frac1{e_1^{6n+3}}
\cr
\Gamma_k&:=\Gamma_1\left(\frac{Ae_1}{d^{4n+3}\delta^{2n}}\right)^{k-1}
\left[
\exp\left(\alpha\frac{k(k-1)}{2}\right)\right]^{6n+3}\ .
}\autoeqno{g.k} 
$$
}

\proof The expressions of $e_1$ and $\Gamma_1$ are a direct
consequence of lemmas \lemmaref{phi} and \lemmaref{delta}. By
induction assume that the estimates of the lemma are true for $k$ we
prove them for $k+1$.
By lemmas \lemmaref{phi} and \lemmaref{delta} we have
$$
\left|r_k^{\indi{k-1}}\circ\phi^{\tau_k}\right|_{(\sigma-k\delta)e_ke^{-\sigma
\tau_k},(\rho-kd)e_ke^{-\sigma
\tau_k} }\leq \Gamma_k
$$
and therefore
$$
\eqalign{
\left|r_{k+1}^{\indi{k}}\right|_{(\sigma-(k+1)\delta)e_ke^{-\sigma
\tau_k},(\rho-(k+1)d)e_ke^{-\sigma
\tau_k} }
\cr
=
\left|\De(r_{k}^{\indi{k-1}}\circ\phi^{\tau_k})
\right|_{(\sigma-(k+1)\delta)e_ke^{-\sigma
\tau_k},(\rho-(k+1)d)e_ke^{-\sigma
\tau_k} }
\cr
\leq \Gamma_k\frac A{(e_ke^{-\alpha t})^{6n+3}d^{4n+3}\delta^{2n}} 
}$$
This yields $e_{k+1}=e_ke^{-\alpha t}$, and therefore \eqref{e.k}; moreover
$$
\Gamma_{k+1}=\Gamma_k\frac A{(e_ke^{-\alpha
t})^{6n+3}d^{4n+3}\delta^{2n}} \ ,
$$
whence
$$
\Gamma_k=\left(\frac A{d^{4n+3}\delta^{2n}}\right)^{k-1}\Gamma_1\left(
\prod_{i=2}^{k}\frac 1{e_i}\right)^{6n+3}\ ,
$$
This proves \eqref{g.k} upon insertion of \eqref{e.k}. 
This proves the lemma.\quadratino
\vskip 0.1cm\noindent
\lemma{o.rk}{For any $N\geq 2$ one has 
$$
\eqalign{
\norma{\op{r_N^{\indi{N-1}}}}_{L^2\to L_2}\leq
\cr
\left[Ee^{7\alpha t}N^{6n+3}\right]^N BFe^{(4n+2)\alpha t}
\left[\exp\left( \alpha\frac{N(N-1)}2t\right)\right]^{6n+3}\ ,
}\autoeqno{o.1}
$$
where $E,F$ are positive constants independent of $N$.
}
\proof We estimate the l.h.s. of \eqref{o.1} by the $L^1$ norm of the
Fourier transform of $r_N$. By lemma\lemmaref{hat} we have, for
$k\in\Re^{2n}$, 
$$
|\hat r_N^{\indi{N-1}}(k)|\leq\frac{\ccon
t\Gamma_N}{(\rho-Nd)^{2n}e_N^{2n}}
\exp\left[-(\sigma-N\delta)e_N|k|\right]\ ,
$$
and therefore
$$
\eqalign{
\norma{\hat r_N^{\indi{N-1}}}_{L^1}&\leq 
\cr
\frac{\ccon
t\Gamma_N}{(\rho-Nd)^{2n}e_N^{2n}}\int
_{\Re^{2n}}\exp\left[-(\sigma-N\delta)e_N|k|\right] dk
&
=\frac{\ccon
t\Gamma_N}{(\rho-Nd)^{2n}e_N^{2n}}\frac{\ccon
{hi}}{(\sigma-N\delta)^{2n}e_{N}^{2n}}\ . 
}
$$
Choosing $\delta=\sigma/2N$ and $d=\rho/2N$, and inserting the
expressions of $e_N$ and $\Gamma_N$ the assertion is proved because, 
(see e.g.[Ro], Corollary II.19)
$$
 \norma{\op{r_N^{\indi{N-1}}}}_{L^2\to L_2}\leq \norma{
\hat{r}_N^{\indi{N-1}}}_{L^1}
 $$
 \quadratino
\vskip 0.1cm\noindent
\lemma{s.n}{For all $N\geq 2$ and $t\geq 0$ the following estimate holds
$$
\norma{S_N}_{L^2\to L_2}\leq
\left[Ee^{7\alpha t}N^{6n+3}t\right]^N \frac{BF}{N!}e^{(4n+2)\alpha t}
\left[\exp\left( \alpha\frac{N(N-1)}2t\right)\right]^{6n+3}\ ,
$$}
\proof One has
$$
\norma{S_N}\leq\sup_{0<\tau_k<t}\norma{\op{r_N^{\indi{N-1}}}}
\int_0^td\tau_1\int_0^{t-\tau_1}d\tau_2\int_0^{t-\tau_2}d\tau_3...
\int_0^{t-\tau_{k-1}}d\tau_k \ ;
$$
the norm of the Weyl quantization of $r_N$ is estimated by the above
lemma. To compute the integral it is convenient to make a change of
variables introducing the new variables $s_1,...,s_N$ defined by
$$
s_1=t-\tau_1\ ,\quad 
s_1+s_2=t-\tau_2 \ ...\quad s_1+s_2+...+s_N=t-\tau_N \ ,
$$
This transforms the integral into 
$$
\int_0^{t}ds_1\int_0^{t-s_1}ds_2\int_0^{t-(s_1+s_2)}ds_3...\int_0^ 
{t-(s_1+s_2+...s_{N-1})}ds_{N}\ .\autoeqno{I.N}
$$
To see this fact it is enough to remark that
$$
\int_{0}^{t-\tau_{N-1}}d\tau_N=\int_{0}^{t-\tau_{N-1}}ds_N=\int_0^
{t-(s_1+s_2+...s_{N-1})}ds_{N}\ ,
$$
and to iterate the argument.
Denote  $I_N$ the integral in \eqref{I.N}. We claim that
$I_N(t)=t^N/N!$. To this end remark that one has 
$$
I_{N+1}(t)=\int_0^{t}I_N(t-s)ds.
$$ 
Hence the assertion is proved by induction and the result is thus
obtained.\quadratino
\vskip 0.1cm\noindent
{\bf Proof of Theorem \lemmaref{1.1}}. It is enough to apply Lemma
2.9.

\vfill\eject\noindent
{\bf References}
\vskip  0.5cm\noindent
[BIZ] G.P.Berman, A.M.Iomin and
G.M.Zaslavsky,  {\it Method of quasiclassical approximation for $c-$
number projections in coherent state basis}, Physica {\bf 4D} (1981), 113-121
\vskip  0.1cm\noindent
[BZ] G.P.Berman and
G.M.Zaslavsky,  {\it Condition of stochasticity in quantum
nonlinear systems}, Physica {\bf 91}A (1978), 450-460
\vskip  0.1cm\noindent
[Ce] L.Cesari, {\it Asymptotic behaviour and stability problems in
ordinary differential equations}, Springer-Verlag 1963
\vskip  0.1cm\noindent
[Ch] B.Chirikov, {\it A universal instability of many dimensional oscillator
systems},  Phys.Re\-ports {\bf 52}, 1979
\vskip  0.1cm\noindent
[Cha] J.Chazarain, {\it Spectre d'un Hamiltonien quantique et
m\'ecanique classique}, Com\-mun. Par\-tial Diff.Equ\-ations {\bf 6}, (1980),
595-655
\vskip  0.1cm\noindent
[Ch-Iz-Sh] B.Chirikov, F.M.Izraeliev, D.L.Shepelyansky, {\it Quantum
instability}, Soviet Sci. Review Sect.C {\bf 2} (1981), 209-253
\vskip  0.1cm\noindent
[Co-Ro1] M.Combescure and D.Robert, {\it Semiclassical spreading of quantum wave
packets and applications near unstable fixed points of the classical flow},
Asymptotic Analysis {\bf 14} (1997), 377-404
\vskip  0.1cm\noindent
[Co-Ro2] M.Combescure and D.Robert, {\it Semiclassical sum rules and generalized
coherent states}, J.Math.Phys. {\bf 36} (1995), 6596-6610
\vskip  0.1cm\noindent
[Fo] G.Folland, {\it Harmonic analysis in phase space}, Princeton University
Press 1988
\vskip  0.1cm\noindent
[Ha] R.Hagedorn, {\it Semiclassical quantum mechanics}: III, Ann.Phys. {\bf 135}
(1981), 58-70;  IV, Ann.Inst.H.Poincar\'e, {\bf 42} (1985), 363-374
\vskip 0.1cm\noindent
[He] E.Heller, {\it Time-de\-pendent approach to semiclassical dynamics},
J.\-Chem.\-Phys.\-{\bf 62} (1975), 1544-1555
\vskip 0.1cm\noindent
[Pe] I.G.Petrovsky, {\it Lectures on Partial Differential Equations}, Dover 1991
\vskip 0.1cm\noindent
[Ro] D.Robert, {\it Autour de l'approximation semiclassique},
Birkh\"auser-Verlag, 1987
\vskip  0.1cm\noindent
[Vo] A.Voros: D\'eveleoppements semiclassiques, Th\`ese d'Etat,
Universit\'e de Paris XI-Orsay, 1977
\vskip  0.1cm\noindent
[Za] G.M.Zaslavsky, {\it Stochasticity in quantum systems}, Phys.Re\-ports {\bf
80}(1981), 157-250

\bye